\documentstyle[preprint,epsfig,aps]{revtex}
\input epsf

\begin{document}
\title{Josephson current through a correlated quantum level: Andreev
states and $\pi$ junction behavior}

\author{E. Vecino, A. Mart\'{\i}n-Rodero and A. Levy Yeyati}

\address{Departamento de F\'\i sica Te\'orica de la Materia Condensada 
C-V, Universidad Aut\'onoma de Madrid, E-28049 Madrid, Spain}

\date{\today}
\maketitle

\begin{abstract}
The Josephson transport and the electronic properties of a quantum 
dot characterized
by a single level coupled to superconducting leads is analyzed.
Different approximations are used and compared: the mean field 
approximation,
the second-order perturbation theory in the Coulomb interaction and
the exact diagonalization in the zero band-width limit. 
The system exhibits a rich behavior as a function of the relevant
parameters. We discuss in detail the conditions for the observation
of $\pi$ junction behavior and the effect of Coulomb interactions
on the Andreev states. 
\end{abstract}

\section{Introduction}

The observation of the Kondo effect in semiconducting quantum dots
\cite{exp-dot1} has opened a new area of research in which electronic
transport through a nanoscale strongly correlated system can be studied
under controlled conditions. More recently, this effect has also been
observed in carbon nanotubes coupled to metallic electrodes, a system
that can behave in many respects as a quantum dot \cite{exp-dot2}.
These type of devices provide an almost
ideal system to test different theoretical predictions.  
In this direction, a great theoretical interest has arisen in connection
with the possibility of replacing the normal electrodes by
superconducting ones. In fact, carbon nanotubes connected to superconducting
leads, that are already being produced, can provide
a physical realization of such a system \cite{schonenberger2}.      

In these type of systems an issue of fundamental relevance is 
the interplay between electron correlation effects and Andreev
reflection processes, which provide the basic mechanism for transport 
between normal and superconducting regions. This interplay has been 
analyzed by several authors for the case of quantum dot coupled to 
a normal and a superconducting electrode \cite{n-dot-s,nskondo}. 
There have also been some works addressing the problem of the 
electron transport mediated by multiple Andreev reflections 
through a resonant level between two superconducting
electrodes \cite{mar-sds}.

When both electrodes are superconducting a more basic question is how
the electron-electron interactions in the dot would affect the Josephson
current. This issue has received considerable attention in recent years 
\cite{ishizaka,arovas1,ambegaokar,arovas2,arovas3}. 
In particular, the discussion
has been centered to a large extent around the appearance of a
$\pi$-junction behavior induced by electron-electron interactions. 
The $\pi$-junction behavior, which consists in a reversal of the sign of
the supercurrent under certain conditions,
was first pointed out by Kulik \cite{kulik}
when analyzing the Josephson tunneling in the presence of
magnetic impurities, and discussed afterwards by several authors 
\cite{shiba,glazman,kivelson}. This issue is also related to the 
interplay between superconductivity and
magnetism. $\pi$-junction behavior has in fact been recently
observed in S-F-S Josephson junctions \cite{exp-sfs}.
  
The existing theoretical works analyzing the Josephson effect in a
single correlated level
coupled to superconducting leads have adopted either a mean
field description \cite{arovas1} or are restricted to some limiting
situations \cite{ishizaka,ambegaokar}. 
In Ref. \cite{ishizaka} the current was obtained to the
lowest order in the tunneling coupling, thereby neglecting the important
physics associated with the Andreev bound states. On the other hand, 
in Refs. \cite{ambegaokar,arovas2} the limit of infinite Coulomb repulsion was
considered. 

At this point we believe that further work is needed to 
understand the physical behavior of this system for a broad parameter
range. This behavior is actually rather complex due to the several 
energy scales involved. In particular, as the Andreev bound states 
play a crucial role in the transport properties, a detailed analysis of 
electron correlation effects on these states seems desirable.
In this paper we present a theoretical study of the Josephson effect
in a single level quantum dot coupled to superconducting electrodes. 
Different approaches are considered. We first 
study the zero band-width limit in which the problem 
can be exactly diagonalized.
Then, as an approximation to the full model we use
a diagrammatic expansion of the self-energy
associated with the Coulomb interaction in which the coupling to the
superconducting leads is taken into account up to infinite order. This
last ingredient is important for a proper description of the Andreev
bound states. 

In section II, we introduce the theoretical model and discuss the 
diagrammatic approximations for the electron self-energy. In section III
we present a simplified version of this model in which the Coulomb
interaction is replaced by an effective exchange field.
This simple model already describes the $\pi$-junction transition and
allows  to understand the behavior of the Andreev states under
a finite magnetization in the dot. Section IV is devoted to the analysis
of the zero band-width limit in which the model becomes 
equivalent to a finite system which can
be solved exactly. The exact solution is then used as a test of the
approximations used for the electron self-energy. It turns out that the
zero band-width limit can describe rather accurately most of the 
properties of the full
model. The results for the
full model are presented in section V. We first discuss two opposite
limiting situations in which the coupling of the dot to the leads is
either small or large compared to the superconducting gap.
As in the zero band-width limit, in the case of weak coupling
we show that the problem can be solved
exactly. We also present some numerical results for the intermediate
regime. The paper is closed by a brief summary of the main conclusions.

\section{Model and theoretical approach}
\label{model-theory}

We describe a small quantum dot connected to
superconducting electrodes by means of a modified single-level 
Anderson model, given by

\begin{equation}
\hat{H} = \hat{H}_L + \hat{H}_R + \sum_{\sigma} \epsilon_0 \hat{n}_{\sigma}
+ U \hat{n}_{\uparrow} \hat{n}_{\downarrow} + \hat{H}_T,
\end{equation}
where $\hat{n}_{\sigma}=\hat{c}^{\dagger}_{0\sigma} \hat{c}_{0\sigma}$,
$\hat{H}_L$ and $\hat{H}_R$ represent the uncoupled 
superconducting leads; $\hat{H}_T = \sum_{k \in L,R;\sigma}
t_{0,k} \hat{c}^{\dagger}_{0\sigma} \hat{c}_{k,\sigma} + h.c.$
describing the coupling between the dot level and the leads. Within
this model the dot is represented by a single spin degenerate level
with a repulsive Coulomb interaction described by the $U$-term in Eq. (1).
We shall assume that the superconducting leads are well described by the
BCS theory with a superconducting gap $\Delta$ and that there is a
fixed superconducting phase difference $\phi = \phi_L - \phi_R$ between
both electrodes. 

The relevant quantities like the current and the spectral 
densities can be expressed 
in terms of non-equilibrium Green functions \cite{Keldysh}. For
the description of the superconducting state it is useful to introduce spinor
field operators \cite{Nambu}, which in a site representation are defined as:

\begin{equation}
\hat{\psi_{i}} = \left(
\begin{array}{c}
c_{i \uparrow} \\ c^{\dagger}_{i \downarrow}
\end{array} \right) \hbox{  ,     } \hat{\psi}^{\dagger}_{i}=
\left(
\begin{array}{cc}
 c^{\dagger}_{i \uparrow} & c_{i \downarrow}
\end{array} \right)
\end{equation}
\noindent
Then, the different correlation functions appearing in the Keldysh formalism
adopt the standard causal form:

\begin{equation}
\hat{G}_{ij}^{\alpha,\beta}(t_{\alpha},t'_{\beta})=-i
< \hat{T}[\hat{\psi}_{i}(t_{\alpha})
\hat{\psi}_{i}^{\dagger}(t'_{\beta})]>
\end{equation}
\noindent
where $\hat{T}$ is the chronological ordering operator along the 
Keldysh time contour. The labels $\alpha$ and $\beta$ refer to the upper
($\alpha \equiv +$)
and lower ($\alpha \equiv -$) branches on this contour. The  
functions $\hat{G}^{+-}_{ij}$, which can be associated
within this formalism with
the electronic non-equilibrium distribution functions,
are given by the (2x2) matrix:

\begin{equation}
\hat{G}^{+-}_{i,j}(t,t^{\prime})= i \left(
\begin{array}{cc}
<c^{\dagger}_{j \uparrow}(t^{\prime}) c_{i \uparrow}(t)>   &
<c_{j \downarrow}(t^{\prime}) c_{i \uparrow}(t)>  \\
<c^{\dagger}_{j \uparrow}(t^{\prime}) c^{\dagger}_{i \downarrow}(t)>  &
<c_{j \downarrow}(t^{\prime}) c^{\dagger}_{i \downarrow}(t)>
\end{array}  \right) .
\end{equation}
\noindent

In terms of the Fourier transform matrix elements of
$\hat{G}^{+-}_{ij}(t,t')$ one can write the charge and the induced  
order parameter on the dot, as well as the Josephson current:

\begin{equation}
<\hat{n}_{\uparrow}> = 
\frac{1}{2 \pi i} \int^{\infty}_{-\infty} d\omega [\hat{G}
^{+-}_{00}(\omega)]_{11},
\end{equation}

\begin{equation}
<c^{\dagger}_{0\uparrow} c^{\dagger}_{0\downarrow}> = 
-\frac{1}{2 \pi i} \int^{\infty}_{-\infty} d\omega [\hat{G}
^{+-}_{00}(\omega)]_{21},
\end{equation}

\begin{equation}
I_{L,R} =\frac{e}{h}  \int^{\infty}_{-\infty} \sum_{k \in L,R}
d\omega \mbox{Tr} \left[ \sigma_z \hat{t}_{0,k} \hat{G}^{+-}_{k,0}
(\omega) -  \sigma_z \hat{t}_{k,0} \hat{G}^{+-}_{0,k}(\omega) \right] ,
\end{equation}
\noindent
where $I_{L,(R)}$ denotes the current between the left (right) lead and
the dot, $\sigma_z$ is the usual Pauli matrix, and $\hat{t}_{0,k}$ is
the hopping matrix in the Nambu representation 

\begin{equation}
\hat{t}_{0,k} = 
\left( \hat{t}_{k,0} \right)^{\dagger} = 
\left( \begin{array}{cc}
t_{0,k}   &  0 \\
0 & -t_{0,k}^* \end{array}  \right) .
\end{equation}

For the zero voltage case the calculation of the different
$\hat{G}^{+-}(\omega)$ is particularly simple because the following
relation holds:

\begin{equation}
\hat{G}^{+-}_{ij}(\omega)=f(\omega)[\hat{G}_{ij}^{a}(\omega)-
\hat{G}_{ij}^{r}(\omega)] ,
\end{equation}

\noindent
where $f(\omega)$ is the Fermi distribution function, and $\hat{G}^{a,r}$ 
are the advanced and retarded Green functions.
Therefore, the relevant  
quantity to be determined is the dot retarded Green function, which in 
a Nambu $2 \times 2$ representation adopts the form

\begin{equation}
\hat{G}_{00}^r(\omega) = \left[ \omega \hat{I} - \epsilon_0 \hat{\sigma}_z -
\hat{\Sigma}^r(\omega) - \hat{\Gamma}_L (\omega) - \hat{\Gamma}_R
(\omega) \right]^{-1},
\label{retarded-GF}
\end{equation} 
where $\hat{\Gamma}_{L,R}$ are the tunneling rates
given $\hat{\Gamma}_{k} = \pi \rho_F \hat{t}_{0,k} \hat{g}^{k}
\hat{t}_{k,0}$, with
$g^{k}_{11} = g^{k}_{22} = - \omega/\sqrt{\Delta^2 - \omega^2}$ and
$g^{k}_{12} = (g^{k}_{21})^* = 
\Delta/\sqrt{\Delta^2 - \omega^2} e^{i\phi_{k}}$ with $k=L,R$ and $\rho_F$ 
is the normal density of states at the Fermi level (the chemical
potential of the superconducting leads is taken as zero). The self-energy
$\hat{\Sigma}^r(\omega)$ takes into account the effect of Coulomb 
interactions.
To the lowest order in $U$ this is given by the Hartree-Fock Bogoliubov
approximation: $(\hat{\Sigma}^r)_{11} = U <\hat{n}_{\downarrow}>$, 
$(\hat{\Sigma}^r)_{22} =
-U <\hat{n}_{\uparrow}>$; 
$(\hat{\Sigma}^r)_{21} = (\hat{\Sigma}^r)^*_{12} = 
U \langle \hat{c}^{\dagger}_{0\uparrow} 
\hat{c}^{\dagger}_{0\downarrow} \rangle$. 
We shall discuss in \ref{ss.correlation} how correlation 
effects beyond this mean field approximation can be included. 

This model has been analyzed within a mean field approximation
in Ref. \cite{arovas1}. 
In that work the mean field solution was simplified by
neglecting the induced order parameter in the dot and imposing
self-consistency only in the dot magnetization. However, the complete
Hartree-Fock Bogoliubov solution requires the self-consistent
determination of both the diagonal and non-diagonal charges in 
the dot. It should be noticed that the self-consistent 
determination of the induced complex order parameter is in principle
necessary in order to ensure current conservation \cite{conservation}.

\subsection{Inclusion of correlation effects}
\label{ss.correlation}

\begin{figure}[h!]
\vspace{0.5cm}
\begin{center}
\includegraphics[width=80mm]{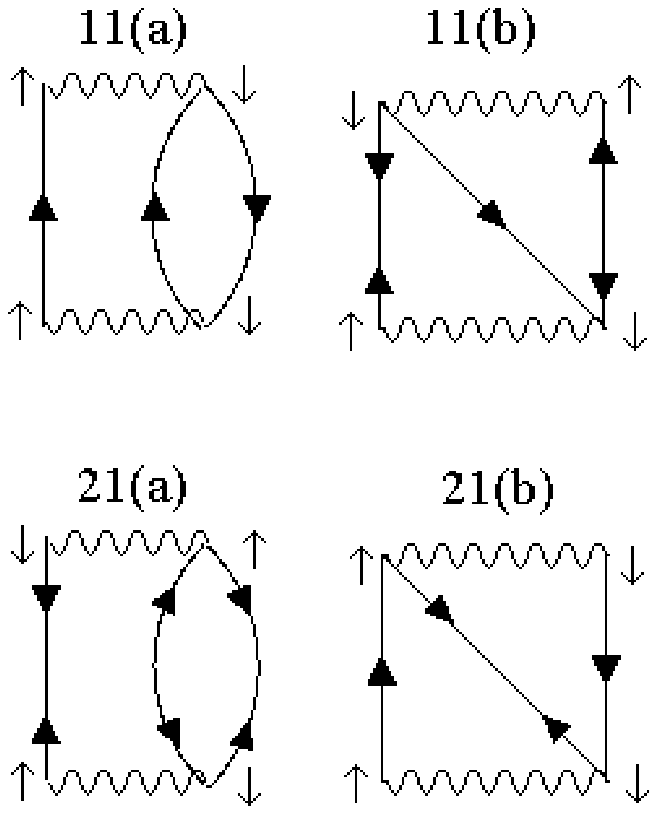}
\end{center}
\caption[]{Diagrams contributing to the second-order self-energy in the
superconducting state.}
\label{diagrams}
\end{figure}

In order to go beyond the mean field approximation, we consider the
diagrammatic expansion of the self-energy in terms of one-electron
propagators in Nambu space. In Fig. \ref{diagrams} we show the 
corresponding second order diagrams in the electron-electron
interaction. Due to the appearance of the anomalous propagators $G_{12}$
and $G_{21}$ in the superconducting state, there are additional diagrams
to the one contributing in the normal state, labeled by 11(a) in Fig. 
\ref{diagrams},
corresponding to the interaction of a quasi-particle with an electron
hole pair with opposite spin.

The proper choice of the unperturbed one-electron Hamitonian over which
the diagrammatic expansion is performed is an important issue. For the
normal Anderson model in the symmetric case $(\epsilon_0 = -U/2)$ the
Hartree approximation which renormalizes the dot level as
$\epsilon_0 + U/2$ is the adequate starting point for the perturbation
theory as it automatically warrants charge consistency between the
perturbed and the unperturbed situation. However, in a
non-symmetric case perturbation over the Hartree field yields 
pathological results close to half-filling (see \cite{friedel}). 
As discussed in Ref. \cite{friedel} a better choice is to define an
effective dot level $\epsilon_{eff}$ in such a way
that charge consistency between the effective and the interacting
problems
is achieved. The natural extension of this scheme to the superconducting
case is to impose also consistency in the non-diagonal charge $n_{12}$
by introducing an effective local pairing potential $\Delta_{eff}$      
in the unperturbed Hamiltonian \cite{nskondo}.
   
With these definitions, the dot Green functions in the unperturbed
effective problem are given by (hereafter we omit the subscript
$00$ in the dot Green functions)

\begin{equation}
\hat{G}^{r(0)}(\omega) = \frac{1}{D(\omega)} \left( \begin{array}{cc}
\omega - \epsilon_{eff} - t_L^2 g^L_{11} - t_R^2 g^R_{11} &
-\Delta_{eff} + t_L^2 g^L_{12} + t_R^2 g^R_{12} \\
-\Delta^*_{eff} + t_L^2 g^L_{21} + t_R^2 g^R_{21} &
\omega + \epsilon_{eff} - t_L^2 g^L_{22} - t_R^2 g^R_{22} \end{array}
\right) ,
\label{effectiveG}
\end{equation}
where $D(\omega)$ is the corresponding determinant.
Notice that the Andreev states in the unperturbed problem are determined
by the condition

\begin{equation}
D(\omega) = 0 ,
\end{equation}
which, in the general case, can have up to four solutions. 

The different contributions to the self-energy represented in Fig. 
\ref{diagrams} can then be written in terms of the
one-electron propagators in the following way

\begin{eqnarray}
\Sigma^{r(2)}(\omega)_{11,a} & = & \frac{U^2}{(2\pi i)^3} 
\int d\epsilon_1 \int d\epsilon_2 \int
d\epsilon_3 \nonumber \\
&& \frac{G^{(0)+-}_{11}(\epsilon_1) G^{(0)+-}_{22}(\epsilon_2)
G^{(0)-+}_{22}(\epsilon_3) \,\, + \,\, G^{(0)-+}_{11}(\epsilon_1)
G^{(0)-+}_{22}(\epsilon_2) G^{(0)+-}_{22}(\epsilon_3)}
{\omega - \epsilon_1 - \epsilon_2 + \epsilon_3 + i0^+} ,
\label{sigma11a}
\end{eqnarray}

\begin{eqnarray}
\Sigma^{r(2)}(\omega)_{11,b} &=& \frac{U^2}{(2\pi i)^3} 
\int d\epsilon_1 \int d\epsilon_2 \int
d\epsilon_3 \nonumber \\
&& \frac{G^{(0)+-}_{12}(\epsilon_1) G^{(0)+-}_{21}(\epsilon_2)
G^{(0)-+}_{22}(\epsilon_3) \,\, + \,\, G^{(0)-+}_{12}(\epsilon_1)
G^{(0)-+}_{21}(\epsilon_2) G^{(0)+-}_{22}(\epsilon_3)}
{\omega - \epsilon_1 - \epsilon_2 + \epsilon_3 + i0^+} ,
\label{sigma11b}
\end{eqnarray}

\begin{eqnarray}
\Sigma^{r(2)}(\omega)_{21,a} &=& - \frac{U^2}{(2\pi i)^3} 
\int d\epsilon_1 \int d\epsilon_2 \int
d\epsilon_3 \nonumber\\
&& \frac{G^{(0)+-}_{21}(\epsilon_1) G^{(0)+-}_{12}(\epsilon_2)
G^{(0)-+}_{21}(\epsilon_3) \,\, + \,\, G^{(0)-+}_{21}(\epsilon_1)
G^{(0)-+}_{12}(\epsilon_2) G^{(0)+-}_{21}(\epsilon_3)}
{\omega - \epsilon_1 - \epsilon_2 + \epsilon_3 + i0^+} ,
\label{sigma21a}
\end{eqnarray}

\begin{eqnarray}
\Sigma^{r(2)}(\omega)_{21,b} &=& \frac{U^2}{(2\pi i)^3} 
\int d\epsilon_1 \int d\epsilon_2 \int
d\epsilon_3 \nonumber \\
&& \frac{G^{(0)+-}_{22}(\epsilon_1) G^{(0)+-}_{11}(\epsilon_2)
G^{(0)-+}_{21}(\epsilon_3) \,\, + \,\, G^{(0)-+}_{22}(\epsilon_1)
G^{(0)-+}_{11}(\epsilon_2) G^{(0)+-}_{21}(\epsilon_3)}
{\omega - \epsilon_1 - \epsilon_2 + \epsilon_3 + i0^+} .
\label{sigma21b}
\end{eqnarray}

Notice that $G^{(0)+-}/2\pi i$ and $G^{(0)-+}/2\pi i$ in the above
equations correspond respectively to the occupied and unoccupied 
states in the dot spectral density. 

The total self-energy is obtained by adding the contributions labeled
by (a) and (b) in the above equations. The remaining self-energy
components ($\Sigma^{r(2)}_{22}$ and $\Sigma^{r(2)}_{12}$) are obtained
by similar expressions. In the non-magnetic case the diagonal self-energy
components are related by $\Sigma^{r(2)}_{22}(\omega) = -
\Sigma^{a(2)}_{11}(-\omega)$. This relation does not hold in the general
magnetic case, except when $\epsilon_0 = -U/2$ for which
$\Sigma^{r(2)}_{11}(\omega) = \Sigma^{r(2)}_{22}(\omega)$.

In order to extend the range of validity of the second order self-energy
an ansatz which interpolates correctly between the limits 
$U/\Gamma \rightarrow 0$ and $U/\Gamma \rightarrow \infty$
can be used \cite{nskondo}.
Within this interpolative approach the matrix self-energy is given by the
following expression \cite{nskondo}

\begin{equation}
\hat{\Sigma}(\omega) = U \langle \hat{n} \rangle \hat{\sigma}_z + \Delta_d 
\hat{\sigma}_x + \left[\hat{I} - \alpha \hat{\Sigma}^{(2)}
\hat{\sigma}_z \right]^{-1} \hat{\Sigma}^{(2)}(\omega) ,
\end{equation}
where 
\[\alpha = \frac{\epsilon_0 + (1-<\hat{n}>)U -\epsilon_{eff}}{U^2
<\hat{n}>(1-<\hat{n}>)}, \]
and $\hat{\Sigma^{(2)}}$ is the second order self-energy. 
 
This interpolative scheme has been used in many different contexts
involving strongly correlated electrons \cite{amr,friedel,kajueter}.
Notice, however, that for the electron-hole symmetric case the
interpolative self-energy reduces to the second order one.

\section{Transition from normal to $\pi$ junction behavior: a simple
model}
\label{simple-model}

An exactly solvable model which can describe the transition
from normal to $\pi$ junction behavior is obtained by substituting the
interacting region by a single site with a local exchange field 
$E_{ex}$, in such a way that $\epsilon_{\uparrow} = \epsilon_0 + E_{ex}$
and $\epsilon_{\downarrow} = \epsilon_0 - E_{ex}$. 
It should be noticed that this model is formally equivalent to a mean
field solution of Hamiltonian (1) with the prescription $E_{ex} = U
(n_{\downarrow} - n_{\uparrow})/2$.  

This model exhibits in general 
four bound states inside the superconducting gap (Andreev states) which give 
the dominant contribution to the current. This is determined by the
derivative of the states below the Fermi energy with respect to the phase. 
The position and spectral
weight of these states can be obtained from the retarded Green function
(Eq. \ref{retarded-GF}). In the limit $\Delta << \Gamma$ and when
$\Gamma_L = \Gamma_R = \Gamma$, the Andreev states can be 
determined analytically by the expression

\begin{eqnarray}
\left(\frac{\omega_{\pm}}{\Delta}\right)^2 =
\frac{\cos^2{\phi/2} + 2 E^2 + Z^2(Z^2 + \sin^2{\phi/2})
\pm 2 X S(\phi)}
{Z^4  + 2 (X^2 + E^2) + 1},
\label{estados}
\end{eqnarray}
where $E = \epsilon_0/2\Gamma$, $X = E_{ex}/2\Gamma$ and $Z^2 =
X^2 - E^2$ and $S(\phi)$ is given by

\[S(\phi) = \sqrt{Z^2 \cos^2{\phi/2} + E^2 + \sin^2{\phi}/4} . \]

It can be
verified that in the limit $X \rightarrow 0$ Eq. (\ref{estados}) 
recovers the well known expression for the Andreev states in a  
single channel contact of transmission $\tau = 1/(1 + E^2)$, 
i.e. $\omega/\Delta = \pm \sqrt{1 - \tau \sin^2{\phi/2}}$
\cite{single-channel}. 

The qualitative behavior of the Andreev states can be easily analyzed in
the particular case $E=0$, in which the position of the four states is
given by 

\begin{equation}
\omega_{\pm}^{\pm} = \pm \Delta
\frac{\cos{\phi/2} \pm X \sqrt{\sin^2{\phi/2} + X^2}}
{(X^2 + 1)}.
\end{equation}

\begin{figure}
\begin{center}
\includegraphics[width=80mm]{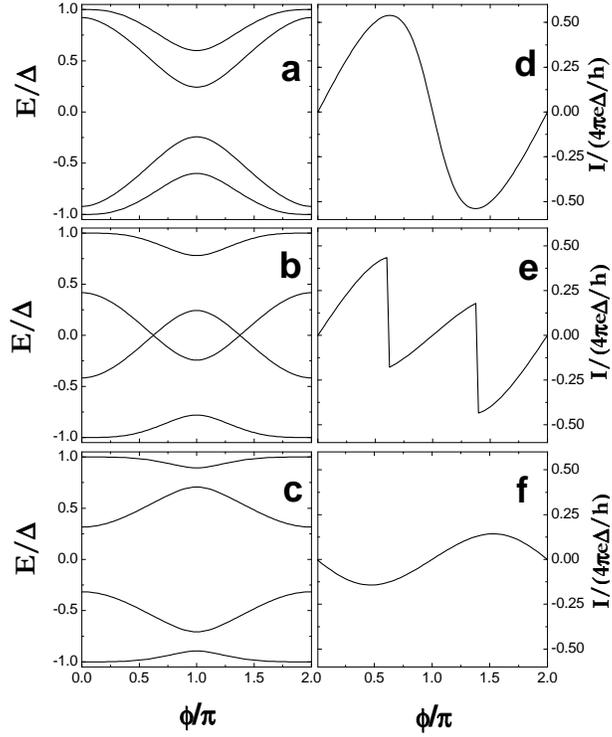}
\end{center}
\caption[]{Bound states within the energy gap and current-phase relation
for the simple model discussed in Sect. \ref{simple-model}. For
$\epsilon_0/2\Gamma = -0.5$ and
$E_{ex}/2\Gamma = 0.25$ (top panel), 0.75 (middle panel), 1.50 (lower
panel).}
\label{sm-cross}
\end{figure}

This expression clearly shows that the effect of a finite exchange field
is to split the non-magnetic Andreev states at $\pm \Delta \cos{\phi/2}$ 
into four states. The transition to the $\pi$ junction behavior is
associated with the progressive inter-crossing of the two "inner"
bound states. This crossing is illustrated in Fig. \ref{sm-cross} for a 
generic
situation. For small exchange field (Fig. \ref{sm-cross} a), 
the inner states do not
cross and the system is in the $0$ state. As $X$ increases, the inner
states cross at $\pi$ and thus the current-phase relation exhibits a
sign change around $\phi = \pi$. Eventually, when the crossing between
the inner states is complete the whole current-phase relation changes
sign and the system is in the $\pi$ state. The state in the intermediate
region is conventionally designed as $0'$ or $\pi'$ depending on
the relative stability of the minima of the inter-crossing states.

\begin{figure}
\begin{center}
\includegraphics[width=70mm]{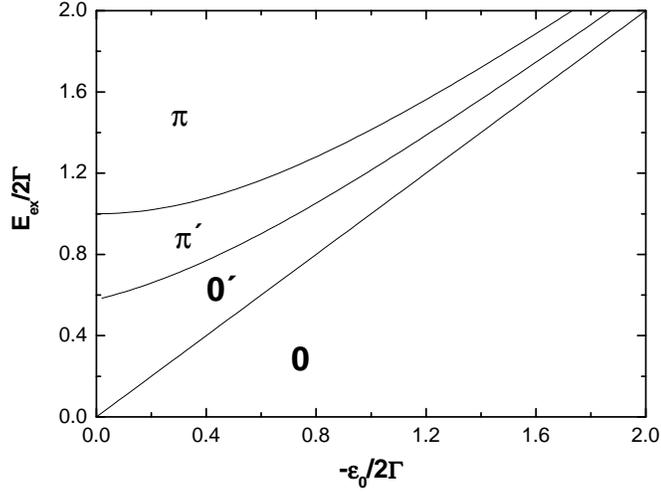}
\end{center}
\caption[]{Phase diagram for the simple model of Sect. 
\ref{simple-model}.}
\label{sm-phasediag}
\end{figure}

The boundaries between the different regions are straightforwardly
obtained from Eq. (\ref{estados}): the curves $X = E$,
$X = (E + \sqrt{3 + 4 E^2})/3$ and $X = \sqrt{1 + E^2}$ correspond
to the $0-0'$, $0'-\pi'$ and $\pi'-\pi$ boundaries respectively. 
The full phase diagram of this simple model is illustrated in Fig.
\ref{sm-phasediag}.

\section{Zero band-width limit}

A special limit in which the model given by Eq. (1) becomes
exactly solvable is the case where the band-width 
of the electrodes tends to zero. This is equivalent to neglect the
high energy excitations in the superconducting electrodes,   
substituting the frequency dependent electrodes self-energy by an
effective diagonal and non-diagonal level in Nambu space \cite{affleck}. 
In this limit the semi-infinite leads connected to the dot are replaced
by an effective single site and the terms $H_{L,R}$ describing the leads 
in Eq. (1) are given by

\begin{equation}
H_{L,R} = \sum_{\sigma} \epsilon_{L,R} 
\hat{c}^{\dagger}_{L,R \sigma} \hat{c}_{L,R \sigma}
\; + \; \Delta_{L,R} \hat{c}_{L,R \uparrow} \hat{c}_{L,R \downarrow} 
\; + \; \Delta^*_{L,R} \hat{c}^{\dagger}_{L,R \downarrow} 
\hat{c}^{\dagger}_{L,R \uparrow}.
\end{equation}

The system thus becomes a sort of ``superconducting
molecule" with a finite number of electron configurations which can be
diagonalized exactly. This solution can be useful as a test of
the accuracy of the proposed approximations for the infinite system.
In spite of its simplicity this limit also gives a qualitative 
description of the behavior of the full model, as will be shown below.

The exact solution in this limit is obtained by
considering electronic configurations with all possible total number
of electrons, the pairing interaction 
connecting configurations that differ in two electrons. 
The eigenstates can be classified into those arising from
even or odd number configurations. In the even case the ground state has zero
total spin, while in the odd case the ground state is two-fold degenerate 
corresponding to $S_z = \pm 1/2$. 

\begin{figure}
\begin{center}
\includegraphics[width=90mm]{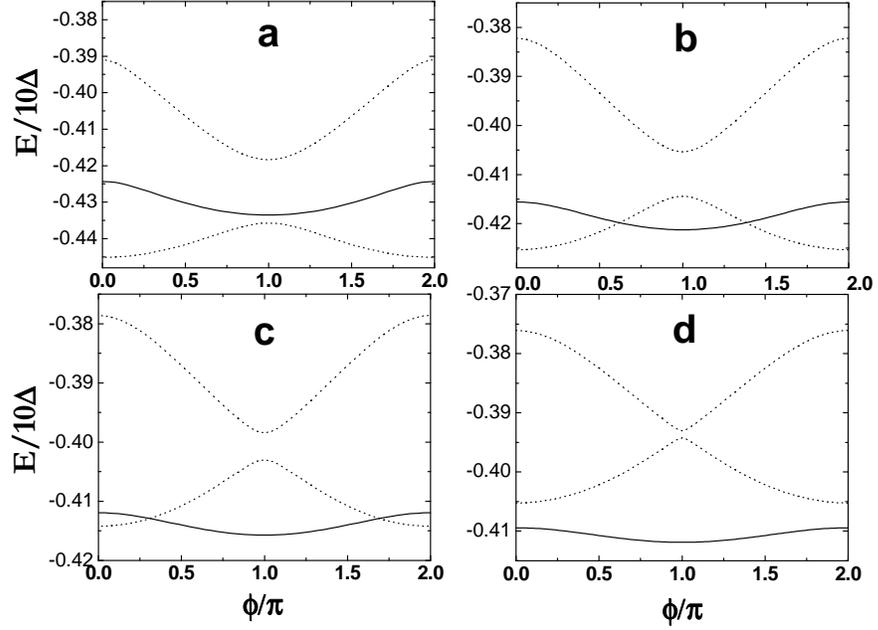}
\end{center}
\caption[]{Energy levels for the model in the zero band-width limit
with $t_L = t_R = 1.2 \Delta$, $\epsilon/\Delta = -10$ and $U/\Delta =
11$ (a), 13 (b), 15 (c) and 18 (d). Dotted and full lines correspond to the
even (ground and first excited states) and odd cases respectively.}
\label{mol-cross}
\end{figure}

In Fig. \ref{mol-cross} we show the evolution of the ground state energy for the
even and odd cases for fixed $\epsilon_0$ and increasing 
$U$. As can be observed, in the even case the ground state energy exhibits
always a minimum for $\phi = 0$, while in the odd case the minimum appears
for $\phi = \pi$. This is to be expected from the fact that the system 
magnetization is finite in the odd case. For small values of $U$ the
even case is more stable for all values of the phase (Fig. \ref{mol-cross}
a), 
while the opposite situation is found for large enough $U$ (Fig. 
\ref{mol-cross} d). 
In the intermediate region (Fig. \ref{mol-cross} b and c) 
the even and odd energy levels cross.
This level crossing corresponds to the
transition between 0 and $\pi$ junction behavior in the full model.
The $\pi$ state thus corresponds to a situation in which the dot
adquires a finite magnetization. 

\begin{figure}
\begin{center}
\includegraphics[width=80mm]{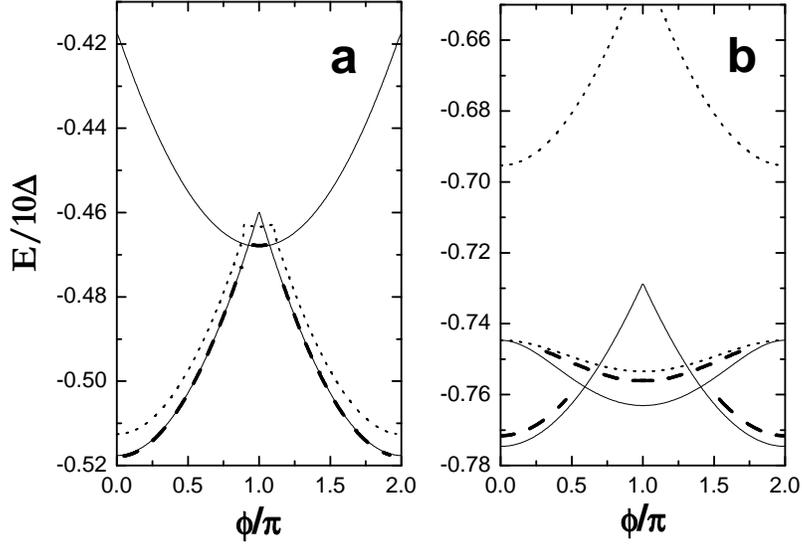}
\end{center}
\caption[]{Comparison of the ground state energy in the zero band-width
limit obtained by different approximations:
exact diagonalization (full lines), mean field approximation (dotted
lines) and second-order self-energy (dashed lines). The values of the
parameters are $\epsilon = -U/2$, $t_L= t_R = 1.2 \Delta$
with $U/\Delta = 2.5 $ (a) and $U/\Delta = 10$ (b).}
\label{mol-comp}
\end{figure}

The exact ground state energy of this simple model can now be used to
check the accuracy of the different approximations discussed in Sect. 
\ref{model-theory}. For the sake of simplicity we shall restrict this
comparison to the $\epsilon_0 = -U/2$ case.  
In Fig. \ref{mol-comp} we show the comparison between the exact, the 
mean field, and the second-order self-energy results
for the ground state energy for different values of $U/\Delta$. At small
values of $U/\Delta$ (Fig. \ref{mol-comp}a), the energy of the mean field 
approximation lies slightly above the exact result both in the non-magnetic
and in the magnetic case. When increasing $U/\Delta$ (Fig.
\ref{mol-comp}b), 
the non-magnetic mean field solution increasingly deviates from the
exact result. In contrast, in the magnetic case, the deviation between
the exact and the mean field solution first increases in the small $U$
range while it progressively decreases for large $U$. This is due to
the fact that the exact ground state corresponds to a localized spin at
the central site for $U \rightarrow \infty$, which is correctly
described by the magnetic mean field solution.

The inclusion of the second-order self-energy substantially improves the
results of the mean field approximation both in the magnetic and in the
non-magnetic cases for small and moderate values of $U$. 
In the non-magnetic case the improvement is considerable in the whole
range of values of $U$ (Fig. \ref{mol-comp}a and \ref{mol-comp}b). 
On the other hand, in the
magnetic case the improvement due to the inclusion of correlation
effects progressively becomes less important as the mean field 
tends to the exact solution in the large $U$ limit.

Finally, it is interesting to analyze the phase-diagram of the model
in the zero band-width limit, which turns out to contain the essential
features of the full model with
the four types of solutions:
0, $0'$, $\pi'$ and $\pi$ appearing in the previous section. 
In this discrete model the different phases can be identified by the
phase-dependence of the ground state energy.

\begin{figure}
\begin{center}
\includegraphics[width=80mm]{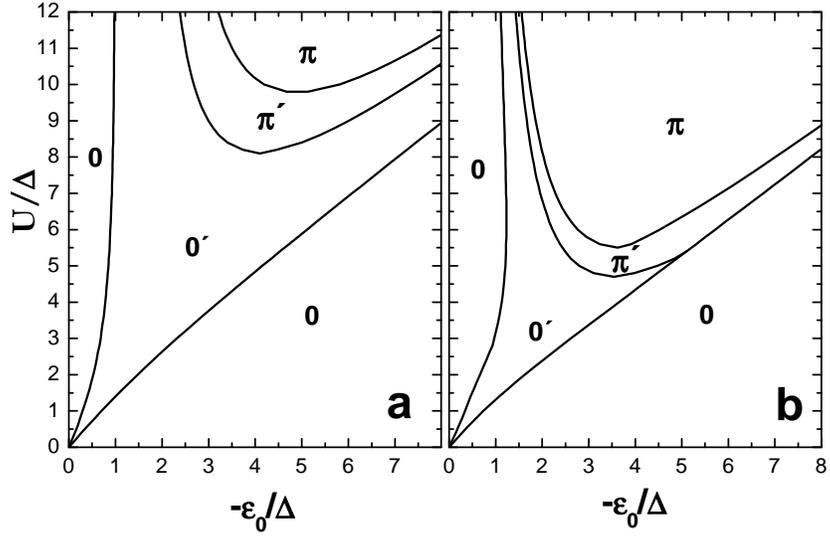}
\end{center}
\caption[]{Phase diagram in the zero band-width limit with $t_L = t_R =
\Delta$ obtained by 
exact diagonalization (a) and within the mean field approximation 
(b).}
\label{mol-phasediag}
\end{figure}

The exact phase diagram for the zero band-width limit is shown in 
Fig. \ref{mol-phasediag}a. 
Only the region $U >0$ and $\epsilon < 0$ is shown as for
all other regions the only possible phase is $0$ type. 
Roughly, the $\pi$ state is found for $U/\Gamma$ and $-\epsilon_0/\Gamma$ 
sufficiently large, while the $0$ state appears either for $U/\Gamma$ or
$-\epsilon_0/\Gamma$ sufficiently small. The $0'$ and the $\pi'$ behavior
is found in the intermediate regime.  It is interesting
to notice that for $-\epsilon_0/\Gamma < 1$ and $U/\Gamma \rightarrow
\infty$ the $0$ state is always the more stable. An extrapolation of
this result to the full model would imply the absence of $\pi$ junction
behavior in the so called mixed valence regime, as has been predicted in
Ref. \cite{ambegaokar}.

For comparison, the mean field diagram is also shown in Fig.
\ref{mol-comp}b. 
Notice that although the overall behavior is
captured in the mean field solution, correlation effects displace the
$\pi$ region to much larger values of $U/\Gamma$ and $-\epsilon_0/\Gamma$.
Neglecting the induced order parameter in the dot as done in
Ref. \cite{arovas1} does not qualitatively change the phase diagram
although the full self-consistent solution
is somewhat closer to the exact result.
  
\section{Results for the full model}

The mean field approximation for the full model has been analyzed in
Ref. \cite{arovas1}. Although in this work self-consistency in the
induced order parameter was neglected, its effects on the total energy
is small as already discussed for the zero band-width limit. In the
present section we will concentrate in discussing the effects of
electronic correlations beyond the mean field solution. We will 
consider the electron-hole symmetric situation ($\epsilon_0 = -U/2$)
in which electron correlation effects are expected to be more important.

For a fixed value of $U$, the physical behavior of the model is
controlled by the dimensionless parameter $\Delta/\Gamma$. Is is
interesting to analyze in detail the two opposite limits $\Delta/\Gamma
\gg 1$ and $\Delta/\Gamma \ll 1$

\underline{$\Delta/\Gamma \gg 1$ limit:} This situation corresponds to
a dot very weakly coupled to the leads. In this limit the exact solution
can be obtained due to the fact that the problem can be mapped into a
two-level Hamiltonian describing the dynamics of the Andreev states.
This can be shown by first considering the non-interacting ($U=0$)
situation. In this case, the spectral density of the dot exhibits
bound states at energies $\omega_s = \pm 2 \Gamma \cos{\phi/2}$ and the
continuous part becomes negligible. Consequently, 
the retarded Green function at the dot can be simply
written as

\begin{equation}
\hat{G}^r(\omega) \rightarrow \left( \begin{array}{cc} \omega &
-\omega_s \\
-\omega_s & \omega \end{array} \right)^{-1} ,
\end{equation} 
indicating that the corresponding effective Hamiltonian 
is given by $\hat{H}^{eff} = \omega_s [\hat{c}_{\uparrow} 
\hat{c}_{\downarrow} + \hat{c}^{\dagger}_{\downarrow} 
\hat{c}^{\dagger}_{\uparrow} ]$. When introducing a finite $U$, 
the diagonal level at $-U/2$ is canceled by the Hartree term and
the remaining part of the Hamiltonian associated with the Coulomb
interaction becomes $U(n_{\uparrow} - 1/2)
(n_{\downarrow} - 1/2) - U/4$. This 
term vanishes for the case of an empty and a doubly occupied dot, while
it yields an energy $-U/2$ for the single electron case. As a
consequence, the ground state of the system is either the symmetric
combination of empty and doubly occupied configurations (with energy
$-\omega_s$) or the doubly degenerate single occupied state (with energy
$-U/2$). Thus, the transition to the magnetic $\pi$ state takes place 
when $U/2 > 2\Gamma$.

\begin{figure}
\begin{center}
\includegraphics[width=100mm]{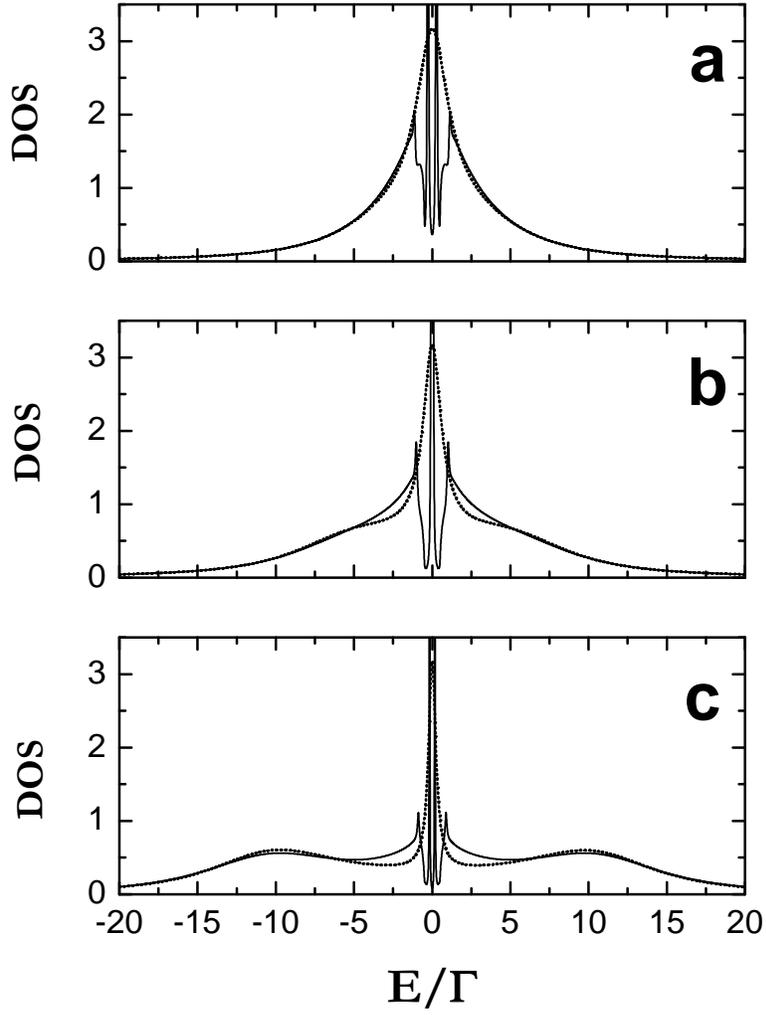}
\end{center}
\caption[]{Density of states for the symmetric case in the regime 
$\Gamma \gg \Delta$. The values of the parameters are $\Delta/\Gamma 
= 0.05$ and $U/\Gamma = 5$ (a), 10 (b) and 20 (c). The dotted lines indicate
the corresponding results for the normal state.}
\label{ldos}
\end{figure}

\underline{$\Delta/\Gamma \ll 1$ limit:} In this opposite limit one
would expect to recover gradually the properties of a normal system. 
In particular, for $U > \Gamma$ the features associated with the Kondo
effect should emerge. Fig. \ref{ldos} shows the dot spectral density in this
regime for increasing values of $U$ obtained in the second-order
self-energy approximation. For comparison the spectral density in the normal
state is also shown. As can be observed, the spectral
density is similar to the one found in the normal state except for the
superimposed features in the gap region. The overall shape evolves as in
the normal case from a single Lorentzian broad resonance for $U <
\Gamma$ to the three peaked structure characteristic of the Kondo regime
when $U > \Gamma$. In this regime the relevant energy scale is set by
the Kondo temperature $T_K$ which essentially measures the width of the
Kondo resonance in the normal state. Within the second-order self-energy
approximation $T_K \sim \Gamma/(1 - \alpha_0)$, where $\alpha_0 =
\frac{\partial \Sigma}{\partial \omega} (0)$. In the symmetric case 

\begin{equation}
\alpha_0 = -\left(\frac{U}{2\pi\Gamma}\right)^2 \left(3- \frac{\pi^2}{4}\right),
\end{equation}
which coincides with the perturbative result of Ref. \cite{yamada} for the
Anderson model. Although this approach fails to give the exponential decay of
the Kondo temperature with $U$ it gives a rather good description of the
spectral density in the moderate $U/\Gamma$ range \cite{Hewson}.

The coexistence of the Kondo and Josephson
effects is to be expected as far as $T_K > \Delta$. When $U$ is further
increased the system should evolved into the magnetic $\pi$ state with 
the suppression of the Kondo effect.

Let us analyze the self-energy in this limit. The effective one-electron
problem in this case is characterized by the presence of Andreev states which, 
from Eq. (\ref{effectiveG}), are located at 
$\omega_s = \Delta \sqrt{1 - \tau \sin^2{\phi/2}}$, where $\tau = 4
\Gamma^2/(\epsilon^2_{eff} + 4 \Gamma^2)$ is the normal transmission 
in the effective problem (in the electron-hole symmetric case
considered here $\epsilon_{eff}=0$ and $\tau=1$), 
like in the case of a single channel point-contact. The weight of the
Andreev states at the dot site decreases as $\Delta/\Gamma$ according to
the expression $\Delta |\sin{\phi/2}|/4\Gamma$. Also the induced order
parameter tends to zero as $\Delta/\Gamma$ in this limit.

The electron self-energy can then be evaluated retaining only
terms of order $\Delta/\Gamma$. The contributions labeled as 11(b) and 21(a) 
can be altogether neglected as they involve more than
one anomalous propagator and are thus of order $(\Delta/\Gamma)^2$.
On the other hand, from the general expressions for the 
second-order self-energy, Eqs. (\ref{sigma11a},\ref{sigma21b}),
it should be noticed that there are
three types of contributions to $\Sigma^{(2)}$: one involving only 
the discrete part of the one-electron spectral densities, another one
involving only the continuous part; and terms in which both the localized
states and the continuous spectrum are mixed. The first of these
contributions is of order $(\Delta/\Gamma)^3$ and can be neglected.
The resulting expression for $\Sigma^{2)}$ up to first order in
$\Delta/\Gamma$ for $|\omega| < \Delta$ is    

\begin{eqnarray}
\Sigma_{11}^{(2)}(\omega) & \simeq &  -\left(\frac{U}{2\pi\Gamma}\right)^2 \left(3-
\frac{\pi^2}{4}\right) \omega
\\
\Sigma_{12}^{(2)}(\omega) & \simeq & \left(\frac{U}{2\pi\Gamma}\right)^2 
\left[|\sin{(\phi/2)}| + 2 (1 +  \frac{1}{\pi}) \cos{(\phi/2)} \right] \Delta .
\end{eqnarray}

\begin{figure}
\begin{center}
\includegraphics[width=90mm]{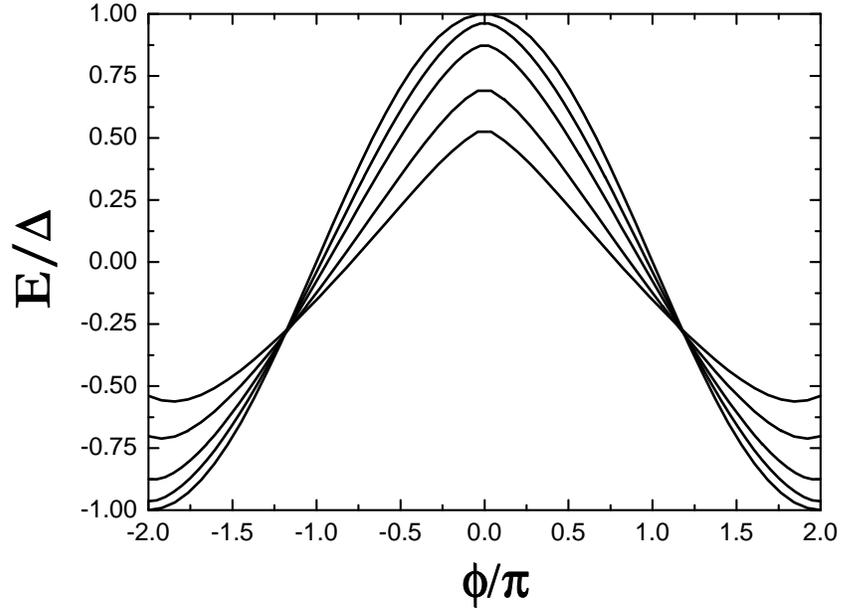}
\end{center}
\caption[]{Renormalized bound states within the gap in the regime
$\Gamma \gg \Delta$ with $U/\Gamma$ = 0, 10, 14, 18 and 20.}
\label{renormalized-states}
\end{figure}

These expressions for the self-energies allow to determine  
the renormalization of the states inside the gap due to the interactions.
For moderate values of $U$ ($U < 10 \Gamma$) the renormalized states
have approximately the same phase-dependence as in the non-interacting
case (i.e. $\sim \cos{(\phi/2)}$) but with a narrower dispersion given by
$\tilde{\omega}_s(0) \simeq \Delta (1 - (U/U_0)^2)$, where
$(U_0/\Gamma)^2 = (\Gamma/\Delta) \pi^2/(2 \pi+ 2)$. 
For larger values of $U$ the phase-dependence of the states
starts to deviate from this simple law. This is illustrated in Fig.
\ref{renormalized-states}. 

The evolution of the renormalized Andreev states indicate that
the critical currents are suppressed as $ \sim (1 - (U/U_0)^2)$ in this
limit (for moderate values of $U$).   

\begin{figure}
\begin{center}
\includegraphics[width=60mm]{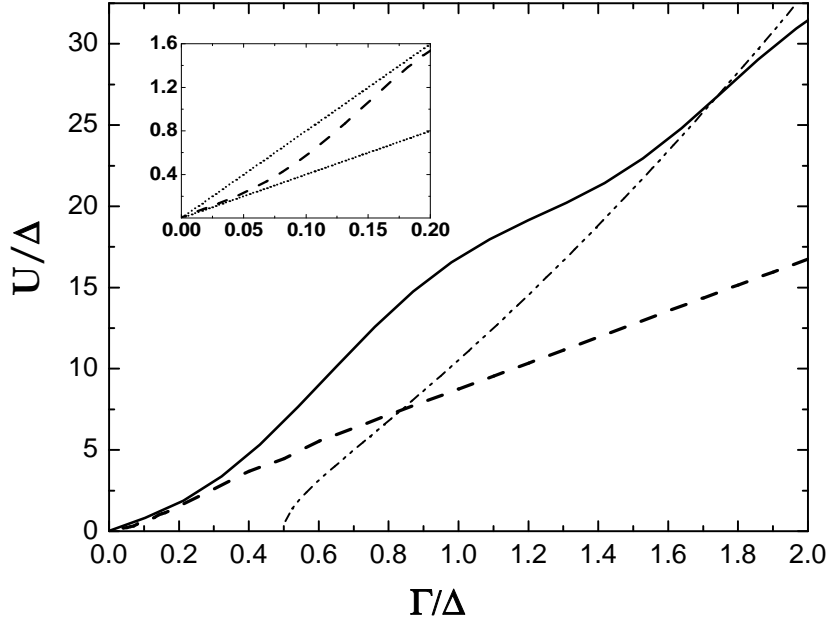}
\end{center}
\caption[]{Phase diagram in the symmetric case for the full model. 
The full and the dashed lines correspond to the onset of the $\pi$
state within the second-order self-energy and the mean field
approximation respectively. The dash-dotted line corresponds to $T_K =
\Delta$ within the second-order self-energy approximation. Inset:
closer view of the $\Gamma \ll \Delta$ region showing the behavior
of the mean field approximation. The dotted lines indicate the
different slopes in the $\Gamma \ll \Delta$ and $\Gamma > \Delta$ regimes.}
\label{sim-phasediag}
\end{figure}

One can summarize the results for the full model in the symmetric case
by discussing the phase diagram shown in Fig. \ref{sim-phasediag}.
In this figure we compare the results of the mean field
and the second-order self-energy approximations for the critical
$U$, $U_c$, defining the transition to the $\pi$ state. 
As can be observed both approximations yield the same result 
in the $\Gamma/\Delta \ll 1$ limit. A close inspection of this regime, 
illustrated by the inset in Fig. \ref{sim-phasediag},
shows that the transition takes place for $U_c \sim 4 \Gamma$ when $\Gamma
\rightarrow 0$ in agreement with the exact result. 
As $\Gamma/\Delta$ increases the predictions
of the two approaches start to deviate. The mean field predicts 
an almost linear relation between $U_c$ and $\Gamma$ although with a larger
slope than in the $\Gamma \ll \Delta$ limit.
On the other hand, the second order
self-energy predicts a faster increase of $U_c$ with $\Gamma$. 
Notice that in the normal state ($\Delta \rightarrow
0$) there should be no transition into a magnetic state for finite
$U$. We also show as a reference the line where $T_K$ in the
normal state, given by the second-order self-energy approximation,
matches $\Delta$. It can be observed that the criterium $T_K \sim \Delta$
for the transition to the $\pi$ state is a rather good for 
$\Gamma$ sufficiently large.

\section{Conclusions and final remarks}

We have presented a theoretical analysis of the Josephson transport through a
strongly correlated level coupled to superconducting electrodes. The analysis
has been specially focused in the effect of electron correlations on the subgap
Andreev states. These states determine to a large extent the 
current-phase relation in this system. The transition to a $\pi$ state can be
understood as a result of the intercrossing of the subgap states induced by 
an increasing Coulomb interaction. Within this model this transition 
corresponds to a truly quantum phase transition in which the ground state
becomes degenerate having a localized magnetic moment at the dot level
as already noticed in Ref. \cite{arovas2}.
It is worth noticing that this situation cannot exist in the absence of 
superconductivity. In fact,
this behavior can be traced to the suppression of the Kondo effect (in 
which the electrons at the dot level couple to a singlet) due to the 
absence of low energy excitations in the superconducting leads.   

In the present analysis we have used different approximation methods. 
In order to get insight on the behavior of the Andreev states in the transition
to the $\pi$ state we have first analyzed a simple mean field model in which
the electron interactions are represented by a local exchange field. 
We have also studied the zero band-width limit which allows for an exact
diagonalization. It has been shown that the study of this limit already
illustrates the different types of behaviors that can be found in the
full model. These results have also been used to determine the accuracy
of the self-energy approach, showing that it improves considerably the
results of the mean field approximation for moderate values of the Coulomb
interaction. Finally, we have presented results for the full model in
different regimes. In the limit $\Delta \gg \Gamma$ we have shown that
the problem can be solved exactly, its dynamics being described by a two
level Hamiltonian corresponding to the Andreev states. On the other
hand, for $\Gamma \gg \Delta$ there is a coexistence of Kondo and
Josephson effects for $T_K > \Delta$, the main effect of electron 
correlations being included in a renormalization of the critical 
current.

The present work constitutes a first step in the study of 
the transport properties of a quantum dot coupled to superconducting
leads in a non-equilibrium situation, i.e. with an applied bias voltage.
This would allow to analyze an experimental situation like the one 
of Ref. \cite{schonenberger2}. Work along this line is under
progress.

\newpage

\newpage

\newpage

\end{document}